\documentclass[epj,nopacs]{svjour}

\usepackage{times}
\usepackage{amsfonts}
\usepackage{amssymb}
\usepackage{amscd}
\usepackage{graphicx}

\newcommand{\be}{\begin{equation}}
\newcommand{\ee}{\end{equation}}
\newcommand{\bea}{\begin{eqnarray}}
\newcommand{\eea}{\end{eqnarray}}
\newcommand{\A}{{\vec A}}
\newcommand{\X}{{\vec X}}
\newcommand{\pro}{\partial}

\newcommand{\V}{{\vec V}}

\newcommand{\B}{{\vec B}}

\newcommand{\nn}{\nonumber}

\newcommand{\hn}{{\hat{\vec n}}}

\begin{document}

\title{Gluonic colour singlets in QCD}
\author{ML Walker}
\institute{Laboratoire de Physique Theorique et Astroparticules, UMR5207, Université Montpellier II, F-34095 Montpellier, France. \email{m.walker@aip.org.au}} 

\abstract{We argue that certain specific gluon excitations, other than conventional 
glueballs, can propagate freely outside of hadrons and glueballs. We begin with the
result that the Abelian generators of $SU(N)$ QCD can be
meaningfully isolated by the Cho-Faddeev-Niemi-Shabanov decomposition. It 
has already been shown by Cho \emph{et. al.} that the corresponding gluon excitations have a gauge transformation whose form indicates that
they are physical entities of neutral colour charge. We contend that this colour 
neutrality permits them to propagate freely, albeit 
with an effective mass due to interaction with the confining 
chromomonopole condensate. According to the dual-superconductor model of the
QCD vacuum, this mass has an upper bound of $k_B T_c \sqrt{2}$. We conclude with some expected
experimental signatures.}
%
\maketitle

\section{Introduction}
Identifying the internal Abelian directions has been of interest to studies of the
QCD vacuum since Savvidy's landmark paper \cite{S77} demonstrating the energetic
favourability of a magnetic condensate. This led to a long-running controversy
surrounding the condensate's stability \cite{NO78,KT00b,H72,S82,DR83,F83b},
with recent papers \cite{F83b,CP02,CmeP04,Cme04,K04,KKP05} concluding in the positive.

What concerns this work is the manner in which the necessarily Abelian internal
direction(s) of the condensate were identified. Two-colour studies typically assigned the Abelian direction to $\hat{e}_3$ \cite{S77,NO78,H72,S82,tH81}, 
in a blatant violation of gauge invariance
that always left doubts that the calculated effects might be gauge artefacts.
A further defect was that these papers were unable
to prove that the magnetic background is due to monopoles.

These problems are avoided by the Cho-Faddeev-Niemi-Shabanov decomposition 
\cite{C80b,FN99b,S99},
which identifies the Abelian directions without choosing a special gauge. It does this by introducing the Cho connection, a topologically generated contribution to 
the gluon field which represents \cite{C80a} a monopole potential. Thus 
the problems of gauge invariance and demonstrating the magnetic condensate to be of
monopole origin are solved simultaneously.

Identifying the Abelian degrees of freedom in a
gauge invariant manner allows us to consider them as physical entities, and
not as gauge artefacts. Furthermore, these particular physical
entities are colour-neutral, and the primary
claim of this paper is that they are not confined. We therefore refer to them
as Free Abelian Gluons (FAGs).

Unconfined gluonic colour-singlets have been discussed in the QCD literature as
glueballs for some time \cite{FM75,MKV09}. However the FAGs discussed in here
are different because they consist of just one gluon, rather than the two or three
confined within gluonium glueballs. Of course, when observed from the outside
they are only distinguished by their masses and, as we shall see in section 
\ref{sec:FAGmass}, their decay modes .

Section \ref{sec:CFNS} presents the CFNS decomposition for general $SU(N)$
gauge groups. Section \ref{sec:unconfined} justifies the claim that the Abelian 
generators are colourless and unconfined, while section \ref{sec:FAGmass} uses the
condensate coupling and dual-superconductor analogy to put an upper limit on 
the FAG's mass, and then
goes on to discuss other properties such as stability and decay modes. Some 
experimental signatures are proposed. 
The paper concludes with a discussion in section \ref{sec:discuss}.

\section{Specifying Abelian Directions} \label{sec:CFNS}
The CFNS decomposition was first presented by Cho \cite{C80a}, and later by 
Faddeev and Niemi \cite{FN99b} and by Shabanov \cite{S99}, 
as a gauge-invariant means of specifying
the Abelian dynamics of two-colour QCD. These authors \cite{C80b,FN99b}
also applied it to three-colour QCD. In this section we adapt it to general
$SU(N)$, although we are not the first to do so \cite{FN99c,LZZ00}, and establish our 
notation. 

The Lie group $SU(N)$ for $N$-colour QCD 
has $N^2-1$ generators $\lambda^{(i)}$, of which $N-1$
are Abelian generators $\Lambda^{(i)}$. For simplicity, we specify the 
gauge transformed Abelian directions with $\hn_i = U^\dagger \Lambda^{(i)} U$. 
Fluctuations in the $\hn_i$ directions are described by $c^{(i)}_\mu$. 
The gauge field of the covariant derivative which leaves
the $\hn_i$ invariant is implicitly defined by
\bea
g\vec{V}_\mu \times \hn_i = -\partial_\mu \hn_i,
\eea
for which the general form is 
\bea
\vec{V}_\mu = c^{(i)}_\mu \hn_i + \vec{B}_\mu ,\qquad
\vec{B}_\mu = g^{-1} \partial_\mu \hn_i \times \hn_i,
\eea
where summation is implied over $i$.

We define the covariant derivative
\be
\hat{\vec{D}}_\mu = \partial_\mu + g\vec{V}_\mu \times .
\ee

It is easily shown that the monopole field strength
\be
\vec{H}_{\mu \nu} = \partial_\mu \vec{B}_\nu - \partial_\nu \vec{B}_\mu
+ g\vec{B}_\mu \times \vec{B}_\nu,
\ee
has only $\hn_i$ components, \textit{ie}. 
\be
H^{(i)}_{\mu\nu}\hn_i = \vec{H}_{\mu\nu},
\ee
where $H^{(i)}_{\mu\nu}$ has the eigenvalue $H^{(i)}$. Since we are only
concerned with magnetic backgrounds, $H^{(i)}$ is considered the magnitude
of a background magnetic field $\vec{H}^{(i)}$.

$\X_\mu$ contains the dynamical degrees of freedom (DOF) perpendicular to 
$\hn_i$, so if $\A_\mu$ is the gluon field then
\bea
\A_\mu &=& \V_\mu + \X_\mu = c^{(i)}_\mu \hn_i + \vec{B}_\mu + \X_\mu,
\eea
where
\bea \label{eq:Xdefn}
\X_\mu &\bot& \hn_i. 
\eea

This appears to leave the gluon field with additional DOF due to $\hn_i, \B_\mu$, 
but detailed analyses can be found in
\cite{CP02,K04,BCK02,KMS05} demonstrating that these fields
are not fundamental, but a compound of dynamic fields.
Hence $\hn_i, \B_\mu$ are dynamic but do not constitute extra DOFs.

Substituting the CFN decomposition into the QCD field strength tensor gives
\bea \label{eq:lagrang}
&\vec{F}^2 =
(\pro_\mu c^{(i)}_\nu - \pro_\nu c^{(i)}_\mu)^2 
+(\pro_\mu \B_\nu - \pro_\nu \B_\mu + g \B_\mu \times \B_\nu)^2& \nn \\
&+ 2(\pro_\mu c^{(i)}_\nu - \pro_\nu c^{(i)}_\mu)\hn_i 
\cdot (\pro_\mu \B_\nu - \pro_\nu \B_\mu + g \B_\mu \times \B_\nu)& \nn \\
&+ (\hat{\vec{D}}_\mu \X_\nu - \hat{\vec{D}}_\nu \X_\mu)^2& \nn \\
&+2g((\pro_\mu c^{(i)}_\nu - \pro_\nu c^{(i)}_\mu)\hn_i 
+ \pro_\mu \B_\nu - \pro_\nu \B_\mu& \nn \\
&+ g \B_\mu \times \B_\nu) 
\cdot (\X_\mu \times \X_\nu)& \nn \\
&+g^2 (\X_\mu \times \X_\nu)^2 
+ 2g (\hat{\vec{D}}_\mu \X_\nu - \hat{\vec{D}}_\nu \X_\mu) \cdot (\X_\mu \times \X_\nu).&\nn \\
\eea
This expression holds for all $N$-colour QCD except $N=2$ where the last term 
vanishes.

The kinetic terms for $c_\mu^{(i)}$ are unmistakably those of Abelian fields.
Eq.~(\ref{eq:lagrang}) has its analogue in studies \cite{S77,NO78,F83b,F80}
utilising the maximal Abelian gauge. However, dependence on a particular gauge
casts a shadow on any analysis and makes it impossible to consider the 
corresponding DOFs as physically significant. However the CFNS decomposition allows
the Abelian dynamics to be specified in a gauge-invariant, well-defined manner
that makes it physically meaningful to say that the fields $c^{(i)}_\mu$ describe
the Abelian component of the gluon field. 

The CFN decomposition also introduces the additional gauge symmetry 
$SU(N)/\left(U(1)^{\otimes (N-1)}\right)$ corresponding to the
gauge transformations of the $\hn_i$, in addition to the original $SU(3)$.
These additional degrees of freedom can be removed by imposing the condition eq.~(\ref{eq:Xdefn}) with the gauge condition \cite{BCK02}
\be \label{eq:newMAG}
\hat{\vec{D}}_\mu \X_\mu = 0.
\ee 
Alternately, one may impose a stronger condition \cite{KMS05}
if Gribov copies are a concern, such as in calculations beyond one loop or  in
lattice studies. Either way, the decomposition is left invariant under the
``active'' \cite{BCK02} gauge transformation
\bea \label{eq:active}
\delta_G c_\mu^{(i)} &=& \hn_i \cdot (\partial_\mu \vec{\alpha}) ,\nn \\
\delta_G \B_\mu &=& (\partial_\mu \vec{\alpha})_{\bot \hn_i}
+ g\B_\mu \times \vec{\alpha} ,\nn \\
\delta_G \X_\mu &=& g\X_\mu \times \vec{\alpha} .
\eea
and we are back to the original $SU(3)$ gauge symmetry as conventional
QCD, \textit{ie.} without the CFNS decomposition, before gauge fixing. One may now
impose a conventional gauge condition, such as Landau gauge \cite{KMS05}, by
fixing the gauge for $c_\mu^{(i)}$ and $\B_\mu$.

A crucial observation of eq.~(\ref{eq:active}) is that $\X_\mu$ transforms like
a source, such as a quark, while the transformation of $c_\mu^{(i)}$ is 
photon-like, consistent with both
colour-neutrality \cite{CP02} and the form of the corresponding kinetic terms in
eq.~(\ref{eq:lagrang}). This interpretation in $SU(3)$ QCD leaves six
off-diagonal, or valence, gluons $\X_\mu$ 
corresponding to the six non-white colour-anticolour combinations, and two
colour-neutral Abelian gluons. Consistent with this
picture, the quark states may be chosen to be eigenvectors of all the Abelian
generators simultaneously, since they commute with each other,
providing a gauge-invariant way to define colour charge \cite{K98b,C00}.


\section{Abelian Gluons are Unconfined} \label{sec:unconfined}
The previous section demonstrated two important things. The first is that the
Abelian components of the gluon field can be meaningfully separated from the
off-diagonal components. The second is that these two different gluon types have
different gauge transformation properties (see eq.~(\ref{eq:active})), which indicate
that the Abelian components are photon-like while the off-diagonal gluons should be 
regarded as coloured sources.

Let us consider this point in the context of the dual-Meissner effect model of confinement.
The coloured sources, quarks and valence gluons, are connected to each other
by flux tubes which bind them into either glueball or hadron states. It follows from
both the infrared Abelian dominance \cite{K98b,C00,diG98,P97,BOT97,EI82} and the dual-Meissner analogy that 
this tube will be filled predominantly with the Abelian gluons, at confinement scales.
However, as we shall now argue, the Abelian gluons themselves are not confined by this mechanism.

Crucially, the Abelian gluons do not carry colour charge, as indicated by
equation (\ref{eq:active}), and nor do they couple to each other, as required for
consistency. This 
follows from the definition of Abelian generator and is easily seen from the structure
constants.

This is not to say that Abelian gluons can travel without restriction. The lagrangian
(\ref{eq:lagrang}) contains the term,
\bea
2(\pro_\mu c^{(i)}_\nu - \pro_\nu c^{(i)}_\mu)\hn_i 
\cdot (\pro_\mu \B_\nu - \pro_\nu \B_\mu + g \B_\mu \times \B_\nu),\nn \\
\eea
which clearly indicates that the monopole background acts like a sink/source
for Abelian fields whose gauge bosons must therefore have an effective mass.
It follows that an Abelian gluon attempting to leave its flux tube must possess enough energy to overcome its mass-gap, as expected by a model based on analogy with conventional magnetic fields
restricted to flux tubes in a type II superconductor, but is then free to propagate throughout space. 

Significantly, there is no corresponding sink/source term for 
the valence gluons, although it has been argued \cite{K04,me07} that a
mass-gap term is generated for them from the interaction term
\bea
(\pro_\mu \B_\nu - \pro_\nu \B_\mu + g \B_\mu \times \B_\nu) 
\cdot (\X_\mu \times \X_\nu).
\eea

\section{The Properties of FAGs} \label{sec:FAGmass}
As discussed above, FAGs are Abelian fields, but massive ones. The mass, 
in principal, is calculable from the properties of QCD. 
Specifically, the FAG mass is
analogous to the photon mass in a superconductor, which varies inversely 
as the London penetration depth. 

As is well-known (\emph{e.g.} \cite{A04}), 
in type II superconductors the London penetration
depth is greater than $\frac{1}{\sqrt{2}}$ times the correlation length, or
\begin{displaymath}
 \lambda > \frac{\xi}{\sqrt{2}}.
\end{displaymath}
This makes the corresponding photon/FAG mass less than $\sqrt{2}$ times the mass 
gap, given by 
\begin{displaymath}
 \Delta = k_B T_c.
\end{displaymath}
According to lattice and other numerical studies \cite{ABDFKKS09}, this lies 
in the energy range 151-193 MeV, giving an upper limit to the FAG mass of 
\begin{equation} \label{eq:FAGmass}
 m_{FAG} < 193 \times \sqrt{2} = 274 \mbox{MeV}.
\end{equation}
This is comparable to the $\pi^0,\pi^\pm$ masses of 135,140 MeV respectively.

Apart from the specific mass value, the FAG has properties very similar
to the $Z_0$, except that it couples only to quarks. Hence it can
couple to a quark-antiquark pair, and from there to a photon and an 
$e^+-e^-$ pair. Another decay mode is into a $\pi^0$, with a photon to conserve
angular momentum. While such decay products are by no means unique, the existence
of FAGs would provide a resonance to these products, providing a signal that would 
be unmistakably stronger at the LHC than at any $e^+-e^-$ collider at the same energy, due to
the quark-only coupling at the bare level.

Another interesting possibility is the interception of a FAG by a virtual pion
emitted by a hadron. The virtual pion could absorb the FAG and use its energy 
to become real while emitting a photon. 
Note that both neutral and charged pions can 
participate in this reaction, so a proton could stimulate a FAG to become a
$\pi^0$ and emit a photon, or to become a 
$\pi^+$ (and emit a photon) while turning itself into a neutron. 

This is quite different from the double-meson decay mode of conventional gluonium.

We end this section with a brief discussion of the internucleon potential.
In principle, FAGs ought to contribute a spin-dependent, short-range, van der Waals
interaction between hadrons. Unfortunately, any FAG signal in polarized 
hadron-hadron scattering will certainly be swamped by meson interactions,
since mesons have approximately half of the mass expected for FAGs, and obscured by 
the uncertainties in the inter-hadron potential as well as by the hadrons' 
complicated spin structure \cite{T08,ABBBB88}.

\section{Discussion} \label{sec:discuss}
We have made a case for non-gluonium gluonic colour singlets. It is based on the
observation that two of the eight gluon generators 
in three-colour QCD are without colour
charge, and that it is colour which is confined. The argument requires the CFNS
decomposition to identify the Abelian directions in a gauge invariant
way. Without this it is impossible to claim that the Abelian degrees
of freedom have real physical meaning. It has been noted that 
the analysis is not sensitive to
the number of colours, with one Abelian degree of freedom in the two-colour case
and $N-1$ of them for $N$-colour QCD.

An attractive feature of the CFNS decomposition, which makes it useful in 
dual-Meissner effect studies, is that it unambiguously identifies the gluon's
monopole degrees of freedom \cite{FN99b,S99,C80a}. It is easily shown furthermore, at least to
one-loop order, that the corresponding monopole condensate is non-zero 
\cite{S77,F80}. Especially
important for this work, a term describing the condensate acting as a sink/source
appears. This not only allows the condensate to restrict the chromoelectric flux
to flux tubes as required by the dual Meissner effect, but also provides a mass gap
for unconfined gluon fluctuations. This mass gap has, according to the dual
superconductor analogy, an upper bound, given by eq.~(\ref{eq:FAGmass}), of
274 MeV.

As mentioned in the introduction, the gluonic colour-singlets
discussed here are conceptually distinct from 
conventional glueballs made up of multiple, bound gluons
(gluonium), despite being experimentally distinguishable only by the decays. 
The ideal FAG consists of one single gluon whose decay modes,
discussed in section \ref{sec:FAGmass}, do not include gluonium's
decay to meson pairs (see \cite{MKV09} and references therein).

Finally, we have identified some experimental signatures of FAGs. Firstly,
as a resonance to $e^+-e^-$ pairs or to photon-$\pi^0$ production which would be 
more apparent at the LHC than at an $e^+-e^-$ detector. Secondly,
the catalysis of pion production by hadrons.

\begin{acknowledgement}
The author wishes to thank S.~Narison, G.~Moultaka and V.~Zhakarov 
for helpful discussions. 
He thanks, for generous hospitality, the physics
department at the University of Montpellier (UM2), the physics department of the University of New South Wales, and the College of Quantum Science at Nihon 
University.
\end{acknowledgement}


\end{document}